# Co-regulation of multi-cellular swirling by individual cellular chirality and boundaries

Xi Li[1], Bin Chen[1,2*]

[1]Department of Engineering Mechanics, Zhejiang University, Hangzhou, People's Republic of China

[2]Key Laboratory of Soft Machines and Smart Devices of Zhejiang Province, Hangzhou, People's Republic of China

*To whom correspondence should be addressed: chenb6@zju.edu.cn

**Abstract**

To understand the relationship between the chirality of individual cells and that of tissues and organisms, we have developed a chiral polarized particle model to investigate the movement of cell populations on substrates. Our model analysis indicates that cells with the same chirality can form distinct chiral patterns on ring-shaped or rectangular substrates. Our model analysis also reveals the importance of coordination between boundary features and individual cellular chirality in regulating the movement of cell populations. This work provides valuable insights into comprehending the intricate connection between the chirality of single cells and that of tissues and organisms.

Chirality is a vital biological characteristic that is highly conserved and plays a critical role in development. The embryonic development of "Spiral Cleavage" is observed across various phyla [1]. Internal organs of vertebrates often demonstrate asymmetry or possess chiral structures [2-4]. Despite extensive research on molecular-level chirality and tissue/organism asymmetry, these two areas have largely been studied independently [5] and the precise connection between them remains unclear [6-8], which highlights the need for the exploration of the chiral movements of cell populations on substrates.

Chiral movements of cell populations on ring-shaped substrates were observed in various in vitro experiments [9], where varied types of cells from different animals, such as mouse myoblast, human stem cell and fibroblast, on ring-shaped substrates displayed distinct clockwise (CW) or counterclockwise (CCW) swirling patterns. On rectangular substrates, polarized fibroblasts with a chiral actin cytoskeleton tended to co-align in an "И" shape, but could transform into an "N" shape when the chiral direction of the actin cytoskeleton was reversed [10]. Currently, the precise mechanisms underlying the emergence of these multi-cellular chiral patterns through cell-cell and cell-substrate interactions, as well as how the individual cellular chirality connects to population chirality, remain poorly understood.

Different theories were developed to understand the mechanisms underlying the movements of cell populations on substrates. Though the cell vertex model [11,12], cellular Potts model [13], active network models [14], continuum models [15,16], et al., were often employed in studies of densely packed cell populations with strong cell-cell adhesion, they may not be suitable for cell populations like myoblasts and fibroblasts, which are generally stiffer and have weak cell-cell adhesions. To describe the movement of such cells, discrete models like the Vicsek model [17], self-propelled particles [18-20], and the connecting cells model [21] may be more appropriate. However, these models may be inadequate in describing the chiral behaviors of cell populations subjected to the boundary constraints.

In the current work, we have developed a chiral polarized particle model to investigate the chiral movement of cell populations on substrates. Consistent with experimental observations, our model successfully predicts the formation of chiral patterns for cell populations on ring-shaped or rectangular substrates. We are also able to identify a critical influence distance of substrate boundaries. With the model, we demonstrate that achieving regular chiral patterns in cell populations requires the coordination between boundary features and cellular chirality. This work provides important insights into the regulation of multi-cellular swirling phenomena on micropatterned substrates.

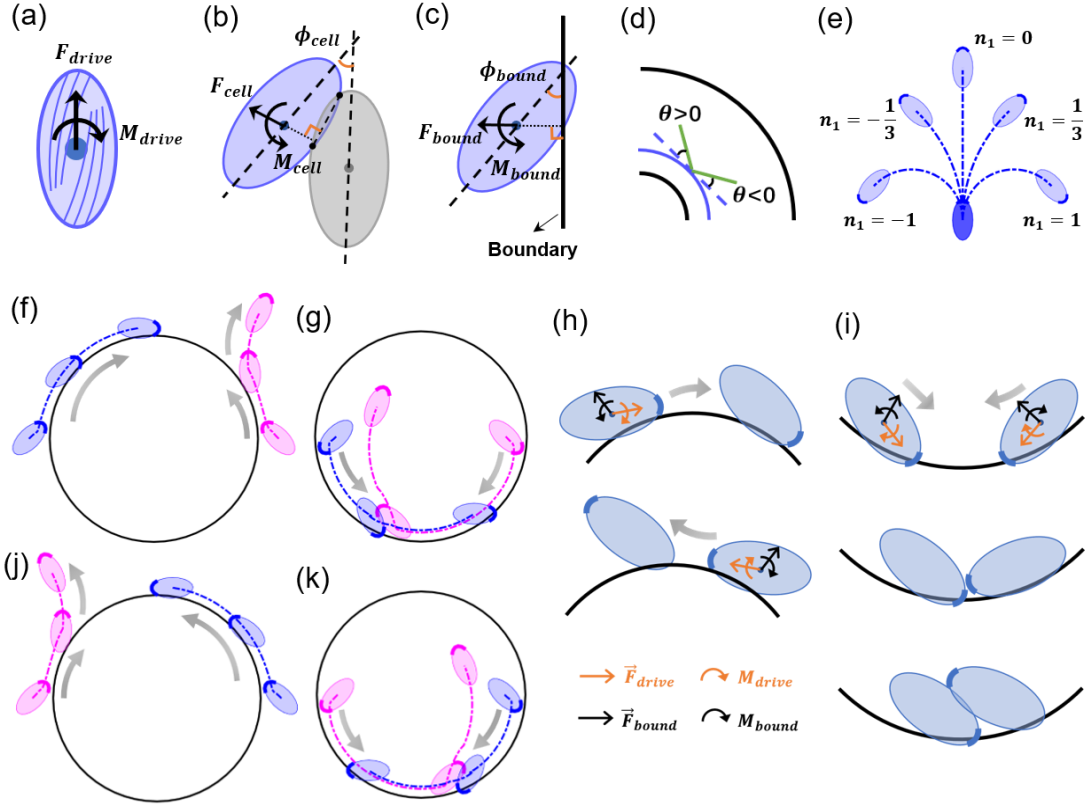

**Fig. 1** (a) Illustration of $F_{drive}$ and $M_{drive}$ due to a chiral cell cytoskeleton; (b, c) Illustration of the force and the moment due to the cell-cell interaction (b) and due to the cell-boundary interaction (c); (d) The definition of the biased angle of cellular alignment, $\theta$; (e) Effect of $n_1$ on the cellular trajectory curvature; (f, g) Movements of individual cells with CW chirality along a convex boundary (f) or a concave boundary (g); (h, i) Force analysis for individual cells with CW chirality along a convex boundary (h) or along a concave boundary (i); (j, k) Movements of individual cells with CCW chirality along a convex boundary (j) or a concave boundary (k).

Generally, the cell needs to be polarized to establish its front and rear in order to move. Chiral cellular movement and alignment was observed in experiments, where Dictyostelium discoideum exhibited a CW migration trajectory on the substrate [22] and human mesenchymal stem cells with chiral actin cytoskeletons attached to stripes displayed biased orientations [23]. It was also observed that fibroblast filopodia curved in one direction [24], filopodia of neuronal growth cones chirally rotated [22,25], and cell cytoskeleton of fibroblasts spontaneously formed a swirling structure on isotropic substrates [10,26]. In view of these experimental observations, we assume in the model that a chiral moment would be induced by a chiral actin cytoskeleton of a cell for its chiral movement, as illustrated in Fig. 1a.

In the model, we initially seed cells to a substrate, which would randomly undergo spontaneous polarization within a short duration. The orientation of cell polarization is also randomly distributed. A polarized cell is simplified as an elliptical particle with the distinct front and rear, which would be subjected to a chiral driving force internally generated by the cell and external forces and/or moments due to the cell-cell interaction and the cell-boundary interaction.

The major axis and the minor axis of a polarized cell are denoted as $2a_1$ and $2a_2$, respectively. The chiral driving force internally generated by the cell is simplified to be equivalent to a concentrated force acting at the cell center along the major axis towards the cellular front, denoted as $F_{drive}$, and a chiral moment, denoted as $M_{drive}$, as illustrated in Fig. 1a. For simplicity, we let both $F_{drive}$ and $M_{drive}$ be constants for a cell, which are related to each other through $M_{drive} = n_1 F_{drive} a_1$, where $n_1$ is a constant, which will be shown to control the direction and strength of the individual cellular chirality. In the model, we further assume $F_{drive}$ to be random and uniformly distributed among cells so that $M_{drive}$ will also be uniformly distributed among cells.

There should generally exist the attractive force and the repulsive force due to the cell-cell interaction. For example, the attractive force may be induced by the cell-cell adhesion, while the repulsive force can result from the elasticity of distorted cell cytoskeletons. In the model, the cell-cell interaction is equivalently simplified as a concentrated force acting at the cell center, denoted as $F_{cell}$, and a moment, denoted as $M_{cell}$. We assume that $F_{cell}$ changes linearly with the intersection area of two neighboring cells and the attraction force is constant so that $F_{cell} = k_{cell} \Delta s_{cell} - F_{cell}^0$ for $\Delta s_{cell} > 0$, where $k_{cell}$ is an elastic coefficient, $\Delta s_{cell}$ is the intersection area of two neighboring cells, and $F_{cell}^0$ is the attraction force between two neighboring cells, which is set to be constant. As illustrated in Fig. 1b, the direction of $F_{cell}$ is set to be perpendicular to the crossline of the two cells.

Note that the cellular polarity was observed to align toward neighboring velocities [27,28], which is similar to nematic liquid crystals [29,30]. Indeed, liquid crystal theory was often employed to investigate the collective cell migration, where the order parameter of liquid crystal was set to be sinusoidally related to orientation of liquid crystals [31]. We then approximate $M_{cell}$ as $M_{cell} = F_{cell} a_{cell} \sin(2\phi_{cell})$ for $\Delta s_{cell} > 0$, where $a_{cell}$ is a constant, which is of a dimension of length, and $\phi_{cell} \in [0,90°]$ is the angle formed by the major axis of two neighboring cells. Note that $F_{cell}$ or $M_{cell}$ vanishes when $\Delta s_{cell} \leq 0$. In our formulation, $M_{cell}$ would tend to drive neighboring cells to be parallel to each other, as illustrated in Fig. 1b.

In the model, the cell-boundary interaction is equivalently simplified as a concentrated force acting at the cell center, denoted as $F_{bound}$, and a moment, denoted as $M_{bound}$. Similar to $F_{cell}$ or $M_{cell}$, we let $F_{bound} = k_{bound} \Delta s_{bound} - F_{bound}^0$ and $M_{bound} = n_3 F_{bound} a_1 \sin(2\phi_{bound})$ for $\Delta s_{bound} > 0$, where $k_{bound}$ is an elastic coefficient, $\Delta s_{bound}$ is the area of a cell outside the boundary, $F_{bound}^0$ is the attraction force between a cell and a boundary, which is set to be a constant, $\phi_{bound} \in [0,90°]$ is the angle between the major axis of a cell and the crossline direction of the cell and a boundary, and $n_3$ is a constant. For simplicity, we set $n_3 = 1$ by default. Note that $F_{bound}$ or $M_{bound}$ vanishes for $\Delta s_{bound} \leq 0$. As illustrated in Fig. 1c, the direction of $F_{bound}$ is set to be perpendicular to the crossline of a cell and a boundary and $M_{bound}$ would tend to drive a cell to be parallel to a boundary.

The resultant force, denoted as $\vec{F}$, or the resultant moment, denoted as $M$, on a cell would be given by $\vec{F} = \vec{F}_{drive} + \vec{F}_{cell} + \vec{F}_{bound}$ and $M = M_{drive} + M_{cell} + M_{bound}$, respectively. $\vec{F}$ is decomposed along its major axis direction, denoted as $F_l$, and along the minor axis direction, denoted as $F_s$. Velocities for cell movement will then be given by $V_l = F_l/\eta_1$, $V_s = F_s/\eta_2$, and $w = M/\eta_r$, where $V_l$ is the translational velocity along the major axis direction, $V_s$ is that along the minor axis direction, $w$ is the angular velocity, and $\eta_1$, $\eta_2$ and $\eta_r$ are three viscous coefficients, with $\eta_1$ or $\eta_2$ being of a unit of $nN \cdot h/\mu m$ and $\eta_r$ being of a unit of $nN \cdot \mu m \cdot h$.

The model described above will be employed to investigate the movement of cell populations on micropatterned substrates. Default values of parameters in the model are provided here. $a_1$=20 $\mu m$ [9], $a_2 = 10\ \mu m$ [9], $a_{cell} = 10\ \mu m$, $F_{dirve} \sim 350 \pm 50\ nN$, $F^0_{cell} = F^0_{bound} = 200\ nN$, $\eta_1 = 10\ nN \cdot h/\mu m$, $\eta_2 = 20\ nN \cdot h/\mu m$, $\eta_r = 10000\ nN \cdot \mu m \cdot h$, $k_{cell} = k_{bound} = 10\ nN/\mu m^2$, $n_1 = 1/3$, and $n_3 = 1$. With default values of $F_{dirve}$ and $\eta_1$, the cellular free-run velocity will be ~35 μm/h, close to the reproted vlaue [9]. With default values of $n_1$ and $\eta_r$, the calculated radius of cellular migration trajectory would be ~150 μm, close to the reported value [22]. With default values of $k_{cell}$ or $k_{bound}$ and $F^0_{cell}$ or $F^0_{bound}$, $F_{cell}$ or $F_{bound}$ can reach zero with $\Delta s_{cell}$ or $\Delta s_{bound}$ being about a few percent of the cellular total area, apparently agreeing with the experiment [9]. In the analyis, the biased angle of cellular alignment, denoted as $\theta$, is defined as positive or negative from the tangential direction of the ring-shaped substrate to the cellular major axis, as illustrated in Fig. 1d.

With the model described above, we firstly carry out parametric studies of $n_1$. In the analysis, we seed only a single cell on a substrate. As displayed in Fig. 1e, our model analysis indicates that, when $n_1 > 0$, the individual cell will possess a CW chirality to rotate CW; when $n_1 < 0$, the individual cell will possess a CCW chirality to rotate CCW. Our analysis also indicates that the cellular trajectory curvature increases as the absolute value of $n_1$ increases. Thus, $n_1$ in our model controls the direction and strength of the cellular chirality.

We then investigate how individual cells interact with boundaries. Results shown in Figs. 1f-i are for individual cells with a CW chirality. In the analysis, a cell may happen to touch a boundary with the front-right or the front-left. As displayed in Fig. 1f, only when the front-right of the cell touches a convex boundary, the cell will maintain a steady CW movement along the boundary, upon which $M_{drive}$ is able to balance with $M_{bound}$ based on the force analysis in Fig. 1h. As illustrated in Fig. 1g, when the cell interacts with a concave boundary, the cell with CCW movement would generally kick out the cell with CW movement when they bump into each other. The force analysis in Fig. 1i indicates that $M_{drive}$ is only able to balance with $M_{bound}$ when touching the boundary with its front-right. Results shown in Figs. 1j-k are for individual cells possessing a CCW chirality. As seen in Fig. 1j, the individual cell with its front-right touching a convex boundary unstably moves CW along the boundary. The cell with its front-left touching a convex boundary will maintain a steady CCW movement along the boundary. As seen in Fig. 1k, when two individual cells moving in opposite directions along the concave boundary bump into each other, the cell with its front-left touching the boundary moving CW would generally kick out the cell with its front-right touching the boundary moving CCW in the analysis. Putting these together, our model analysis indicates that cells with the CW/CCW chirality stably move with their front-right/front-left touching the boundary and cells with the CW/CCW chirality prefer to move CW/CCW along the convex boundary and CCW/CW along the concave boundary on a ring-shaped substrate.

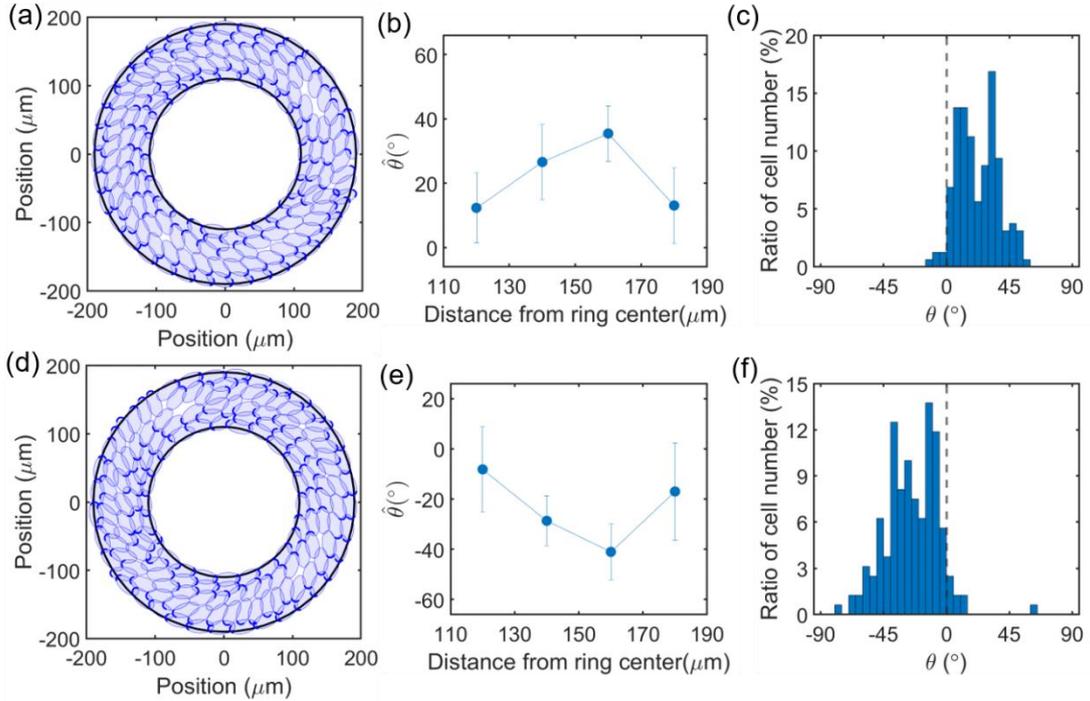

**Fig. 2** Results for the movement and alignment of cell populations with $n_1 = 1/3$ (a-c) and with $n_1 = -1/3$ (d-f) on a ring-shaped substrate: the chiral pattern (a, d), the averaged biased angle at different distance from the center of the ring-shaped substrate (b, e), and the histogram shows the distribution of biased angles of all cells (c, f).

We then investigate the movement of cell population on a ring-shaped substrate. As displayed in Figs. 2a-c, when all individual cells possess the CW chirality with $n_1 = 1/3$, the analysis indicates that cells tend to move CW along the inner boundary and CCW along the outer boundary of the substrate. Cell population demonstrates a CW alignment, characterized by positive biased angles that are smaller near boundaries, but larger in the middle of the ring. In comparison, in Figs. 2d-f, when all cells are set to have a CCW chirality with $n_1 = -1/3$ in the analysis, the directions of cellular movement along boundaries are just reversed and cell population demonstrates a CCW alignment on the substrate.

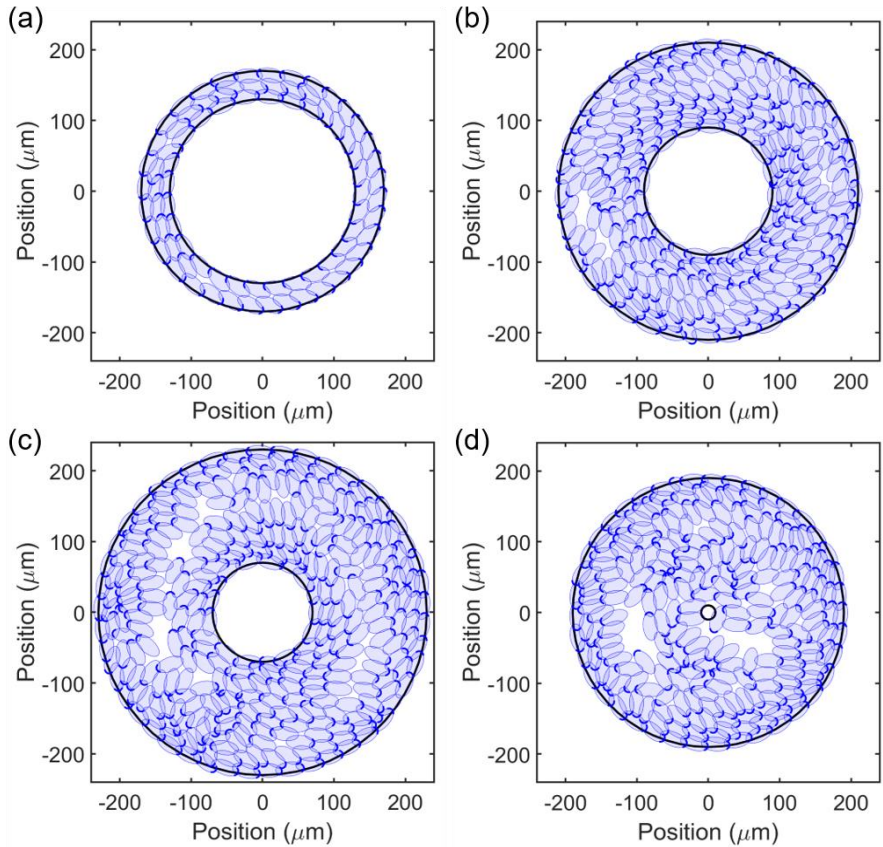

**Fig. 3** Boundary effect on the movement of cell populations on a ring-shaped substrate: (a) The chiral pattern is strong on a narrow substrate; (b-c) On wider substrates, cells near the inner or the outer boundary have regular alignment while cells relatively far from the boundary align irregularly; (d) When the inner boundary is extremely small,

only cells near the outer boundary are regularly aligned while the rest appear to be chaotic. In the analysis, $n_1 = 1/3$.

To furtherly understand the boundary effect on the chiral pattern of cell populations on substrates, we adjust the thickness of ring-shaped substrates in the analysis. As displayed in Fig. 3, decreasing the thickness would make the chiral pattern stronger, while increasing the thickness would weaken the overall chiral pattern. It can be inferred from Figs. 3b-c that both the inner and outer boundaries have a strong influence size about a couple of cells. As displayed in Fig. 3d, when the radius of the inner boundary is too small, the cell population on the substrate is unable to exhibit regular chiral patterns. These results indicate that strong boundary influence is limited within the size of a couple of cells and the regular chiral pattern of cell population exists only on relatively thin ring-shape substrates.

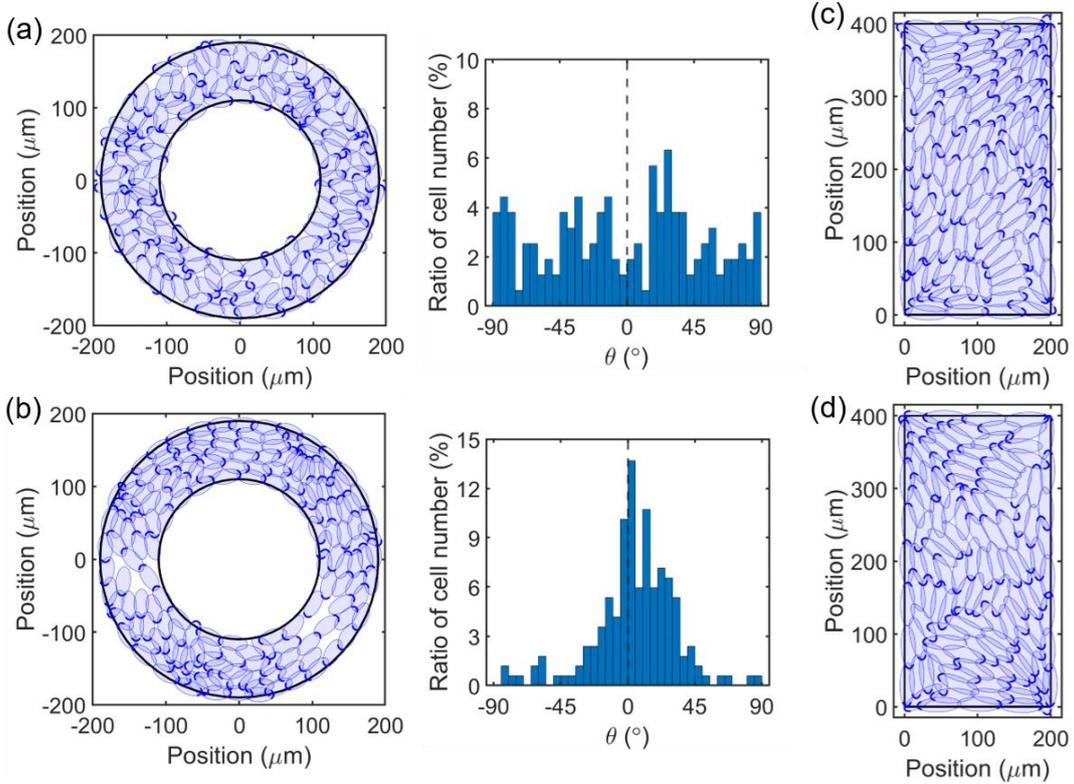

**Fig. 4** (a, b) The chiral pattern of cell population on substrates is irregular when the chirality of individual cells is too strong ($n_1 = 3$) (a) or too weak ($n_1 = 1/100$) (b);

(c, d) On a rectangular substrate, the cell population appears in the "И" shaped arrangement with $n_1 = 1/6$ and transforms into "N" shape with $n_1 = -1/6$ (d). In the analysis for a rectangular substrate, $a_1 = 30\mu m$, $F_{drive} = 200 \sim 220 pN$, $F_{cell}^0 = 100 pN$, and $a_{cell} = 6\mu m$.

To see how the chiral pattern of cell populations on ring-shaped substrates is affected by the strength of individual cellular chirality, we adjust $n_1$ in the analysis. Figure 4a illustrates the chaotic movement of the cell population when $n_1 = 3$, where the cells cannot move stably along the boundaries of a ring-shaped substrate. As shown in Fig. 4b, when $n_1 = 1/100$, jamming occurs and the proportion of positive and negative biased angles is about the same, which also prevents the formation of a chiral pattern. Comparing these results with the analysis for $n_1 = 1/3$ in Figs. 2a-c suggests that the strength of single cellular chirality should fall within an appropriate range in order to generate regular chiral patterns of cell populations on ring-shaped substrates.

We also study the chiral arrangement of cell population on a rectangular substrate with our model, as shown in Figs. 4c, d. In Fig. 4c, when all individual cells have CW chirality, the cell populations form the "И" shape arrangement on the rectangular substrate and the biased angles of most cells are positive. In Fig. 4d, when the individual cellular chirality is reversed to CCW, the cell populations form the "N" shape arrangement on the rectangular substrate and the biased angles of most cells becomes negative. These predictions appear to agree with experiments [10] very well.

Inspired by hydrophilic and hydrophobic boundaries existing in nature [32], we suspect that there might also exist the situation, where the major axis of cells tend to be vertically aligned along a boundary. We can realize such a situation, for example, by setting $n_3 = -0.5$ in the model analysis. Our analysis indicates that, an individual cell with CW/CCW chirality tends to move CCW/CW along a convex "vertical" boundary and CW/CCW along a concave "vertical" boundary, which is exactly the opposite to predictions for "parallel" boundaries. When the outer boundary is designed as a "vertical" boundary and the inner boundary as a "parallel" boundary on a ring-shaped substrate, we find that the movements of cell populations with CW chirality along both boundaries are CW. The averaged biased angle of cells changes from ~0° to ~40° and then gradually to ~-70° in the radial direction, which, interestingly, bears significant similarities to the change of myocardial fiber orientation angle along the ventricular wall thickness [33,34].

As seen from above, our model makes interesting predictions about chiral movements and alignments of cell populations on substrates, which are largely consistent with experiments. However, we must acknowledge that this model may be overly simplified in different ways. For instance, there may be specific intercellular interactions among cells [35-38], which have not been explicitly considered in our model. Furthermore, complex protein structure variations and intricate biochemical signal transmission can be involved, which are not accounted for in our model. Despite these limitations, our model has shown promising results and has provided valuable insights into the movement of cell populations on micropatterned substrates.

In summary, we have developed a chiral polarized particle model to explore the chiral movement and alignment of cell populations on micropatterned substrates. Our analysis has demonstrated that cells with the same chirality exhibit unique chiral patterns on ring-shaped or rectangular substrates. Furthermore, we have identified an important distance at which the boundaries of the substrate significantly affect the multi-cellular chirality. This underscores the significance of boundary effects on the overall chiral pattern. Our analysis emphasizes the coordination between boundary features and individual cell chirality in regulating the chiral pattern of cell populations on substrates. We believe that this research can offer valuable insights into comprehending the intricate relationship between the chirality of single cells and that of tissues and organisms.

**Acknowledgements**

This work was supported by Zhejiang Provincial Natural Science Foundation of China (Grant No.: LZ23A020004) and the National Natural Science Foundation of China (Grant No.: 11872334).